\documentclass{chi2009}

\usepackage{times}
\usepackage{url}
\usepackage{graphics}
\usepackage{color}
\usepackage[pdftex]{hyperref}

\hypersetup{%
pdftitle={Your Title},
pdfauthor={Your Authors},
pdfkeywords={your keywords},
bookmarksnumbered,
colorlinks,
citecolor=black,
filecolor=black,
linkcolor=black,
urlcolor=black,
breaklinks=true,
}
\newcommand{\comment}[1]{}
\definecolor{Orange}{rgb}{1,0.5,0}

\usepackage{amsmath}
\begin{document}

\setlength{\paperheight}{11in}
\setlength{\paperwidth}{8.5in}
\setlength{\pdfpageheight}{\paperheight}
\setlength{\pdfpagewidth}{\paperwidth}


\title{Decoding the Text Encoding}
\numberofauthors{1}
\author{
  \alignauthor Fereshteh Sadeghi and Hamid Izadinia\\
    \affaddr{Computer Science and Engineering}\\
    \affaddr{University of Washington}\\
    \email{\{fsadeghi,izadinia\}@cs.washington.edu}
}

\maketitle

\begin{abstract}
  Word clouds and text visualization is one of the recent most popular and widely used types of visualizations. Despite the attractiveness and simplicity of producing word clouds, they do not provide a thorough visualization for the distribution of the underlying data. Therefore, it is important to redesign word clouds for improving their design choices and to be able to do further statistical analysis on data. In this paper we have proposed a fully automatic redesigning algorithm for word cloud visualization. Our proposed method is able to decode an input word cloud visualization and provides the raw data in the form of a list of (word, value) pairs. To the best of our knowledge our work is the first attempt to extract raw data from word cloud visualization. We have tested our proposed method both qualitatively and quantitatively. The results of our experiments show that our algorithm is able to extract the words and their weights effectively with considerable low error rate.
\end{abstract}



\section{Introduction}
Recently, the use of text visualization and word clouds has become very popular for visualizing various types of data~\footnote{Throughout this document the words \emph{word cloud} and \emph{text visualization} are used interchangeably}. Different tools for generating word clouds from the text data are developed (e.g. Wordle~\footnote{\url{http://www.wordle.net}}) which help understanding the greater prominence to words that appear more frequently in the source text. Of course, the use of word clouds is not limited to text documents and basically each word cloud can provide a visualization of the weights of a list of elements/factors. The word cloud applications usually provide easy to use interfaces for selecting different fonts, layouts, and color schemes and give the user the option to choose the style that is most appropriate for his/her purpose. The word clouds are visually stimulating and easy to digest. The strength of the text visualization lies in the fact that it clearly shows the highlights and 
can convey the high weight words very 
quickly. However, the words with smaller weights can be simply ignored for further comparisons. The text visualization embeds weights of the words and the viewer does not have access to the actual weights and numbers. Therefore, the viewer analysis of the text visualization is only based on the estimation of size which is not exact and is prone to human size estimation bias. In addition, the viewer comparison for the weights of the words is also biased by the choice of colors. According to these factors, we believe that it is needed to have a text visualization decoding tool that can extract the raw data out of the text visualization images.

\begin{figure}[t]
  \begin{center}
  \includegraphics[scale=0.25]{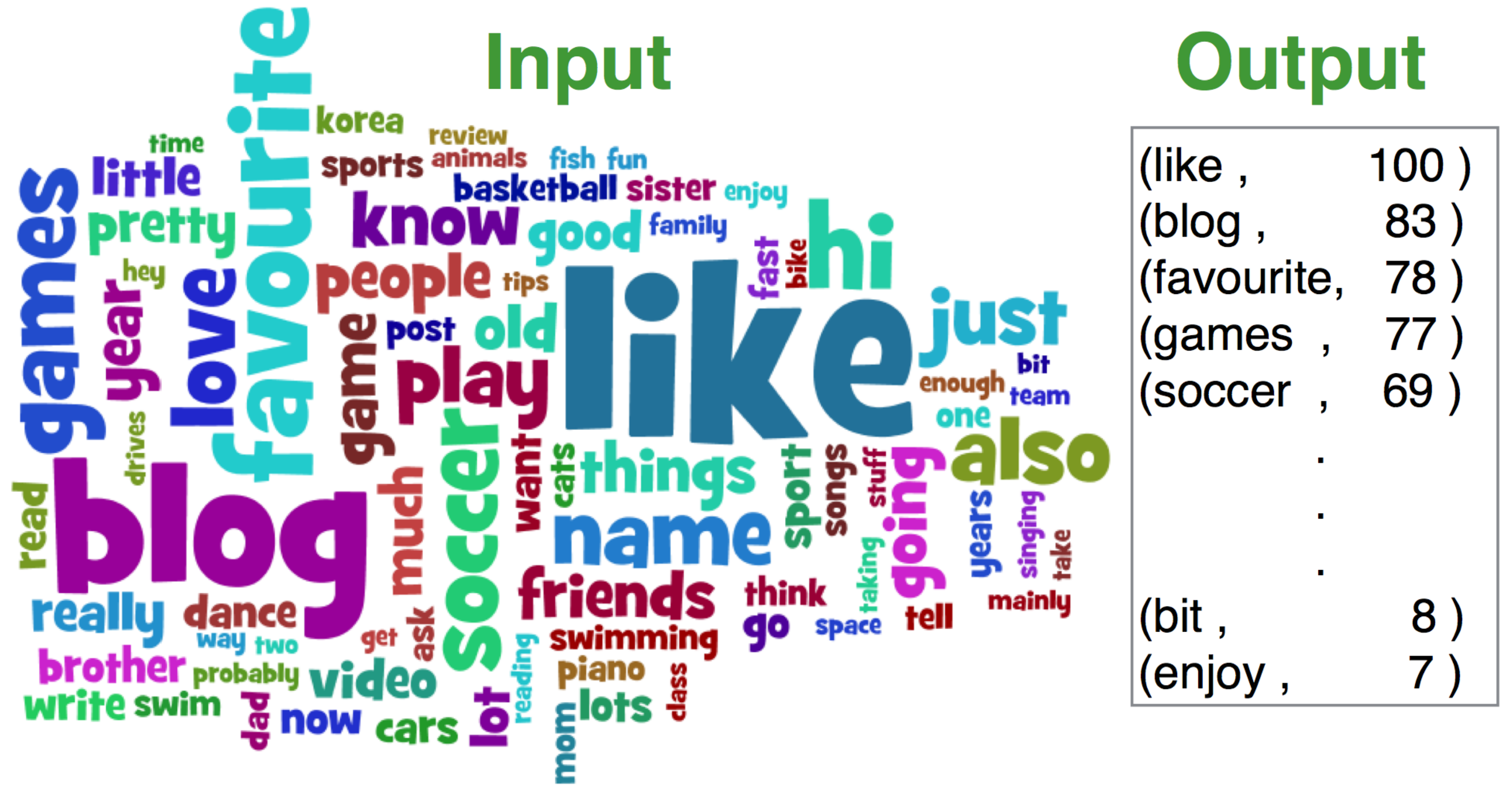}
  \caption{{Our algorithm takes a word cloud as an input image and extracts its raw data in the form of a list of words with their corresponding weights.}
  \label{fig:teaser}}
  \end{center}
\end{figure} 

In this paper, we want to automatically decode and analyze the text visualization diagrams. Given an input bitmap image of a text visualization, we want to automatically extract the keywords along with their weights so that we will be able to automatically extract the raw data information and redesign the input text visualization. This problem is important from two aspects:
\begin{itemize}
\item The use of text visualization, may be a poor design choice in particular applications. Therefore it may hamper understanding of the underlying data which will eventually lead to biased decisions. Extracting the raw data from these visualizations can help us produce better visualizations.
\item The text visualization, provides a diagram which helps the user to compare the weight of different factors ( displayed as words). So, it mainly visualizes two dimensional data in the form of (words,weights). However, the text visualization does not provide any other information about the distribution of data (e.g. the variance, order, etc) which may be useful for further analyzing the data. Extracting the raw weights, will let us compute other statistical information as well as using other methods for visualizing them. 
\end{itemize}
~

\section{Related Work}
In \cite{jeffcolor}, a method for automatic selection of colors is proposed. Previously, \cite{ZhouTan} applied Hough transform to extract bars from bar charts. \cite{Huang} produced vectorized edge maps and applied a set of rules to extract marks from different types of diagrams (bar, pie, line and low-high charts). This technique was further used in a human interaction system for correcting the automatically generated charts~\cite{yang}. 
In the most recent attempts for automatically analyzing and improving the poor design of charts, \cite{ReVision} proposed a method that interprets the bar charts and pie charts using the computer vision techniques and the actual numerical information is extracted using optical character recognition (OCR). Our work, is also aimed to extract raw data from available visualizations however, we are different from \cite{ReVision} in two main aspects: 

\begin{itemize}
\item In \cite{ReVision}, considerable amount of work is done for classifying the different types of charts based on low level visual features. Here, our focus is to extract information specifically from text visualizations since we believe that using state-of-the-art computer vision techniques, the discrimination of text visualization diagrams from other types of diagrams is trivial. 
\item Our underlying mathematics and algorithm for extracting the raw data is totally different from the ideas and procedures used in \cite{ReVision}. The main idea used in \cite{ReVision} is based on several rules specific to pie charts and bar charts (e.g. bar charts have horizontal lines and vertical rectangles which can be extracted using simple shape matching and segmentation. Also, pie charts have several sectors with different colors which can be extracted based on color segmentation)
Our work is based on visual extraction of homogeneous elements (visual letters), detecting which alphabet these letters refer to using OCR techniques and finally building a graphical model on top of the extracted letters to build words. 

\end{itemize}


In all of the aforementioned works the aim is to redesign and improve the diagrams using the available information in the diagrams. 
Our method will extract the \emph{missing} weight values using only visual clues. The result of our work can be used both in redesigning the text visualization diagram as well as extracting the \emph{missing} raw information.

\section{Our Approach}
\begin{figure}[t]
  \begin{center}
  \includegraphics[scale=0.17]{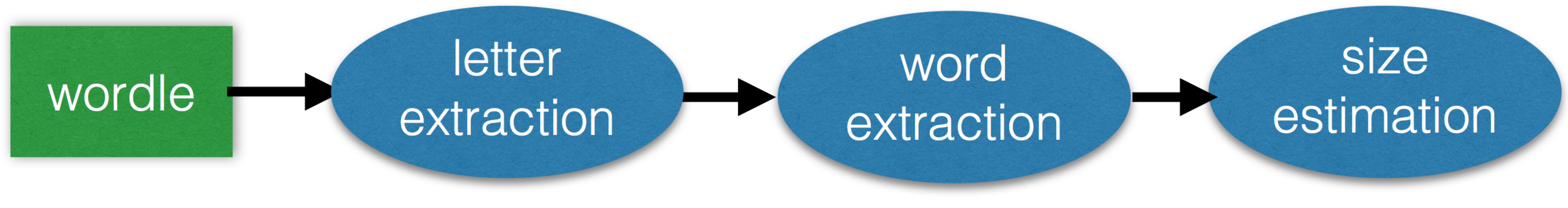}
  \caption{{Our data extraction algorithm has three main steps of letter extraction, word extraction and size estimation which are executed in a pipeline}
  \label{fig:diagram}}
  \end{center} 
\end{figure}

For extracting the raw data of a word cloud we have designed an algorithm with three main steps which is summarized in Fig.~\ref{fig:diagram}.
Here we explain the major steps of building our system. The input of our system will be a bitmap image of the text visualization and the output will be a list of (nominal variable, value) pairs.

\begin{figure}[t]
  \begin{center}
  \begin{tabular}{c}
  \includegraphics[scale=0.2]{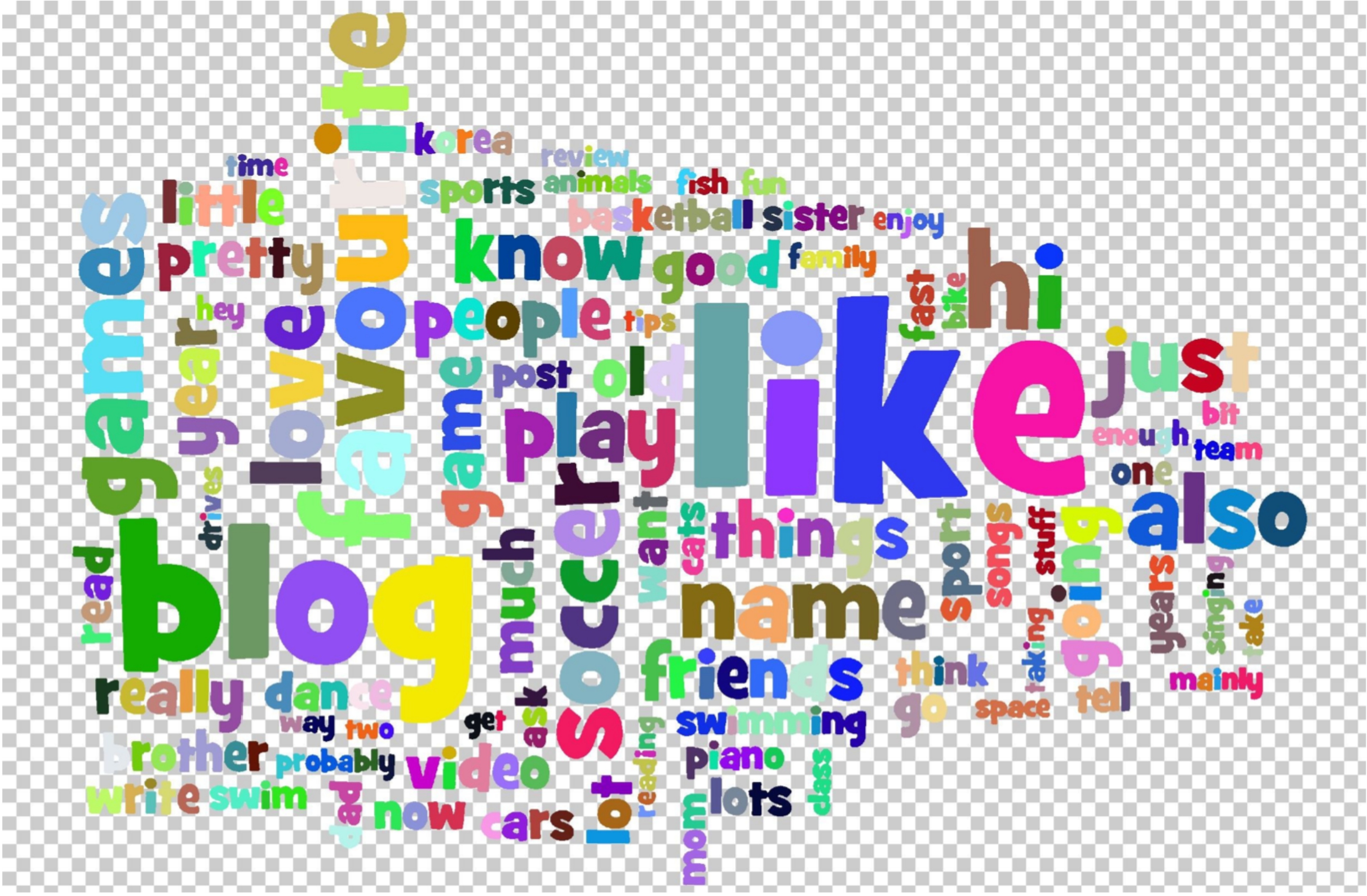} \\
  (a) \\
  \includegraphics[scale=0.2]{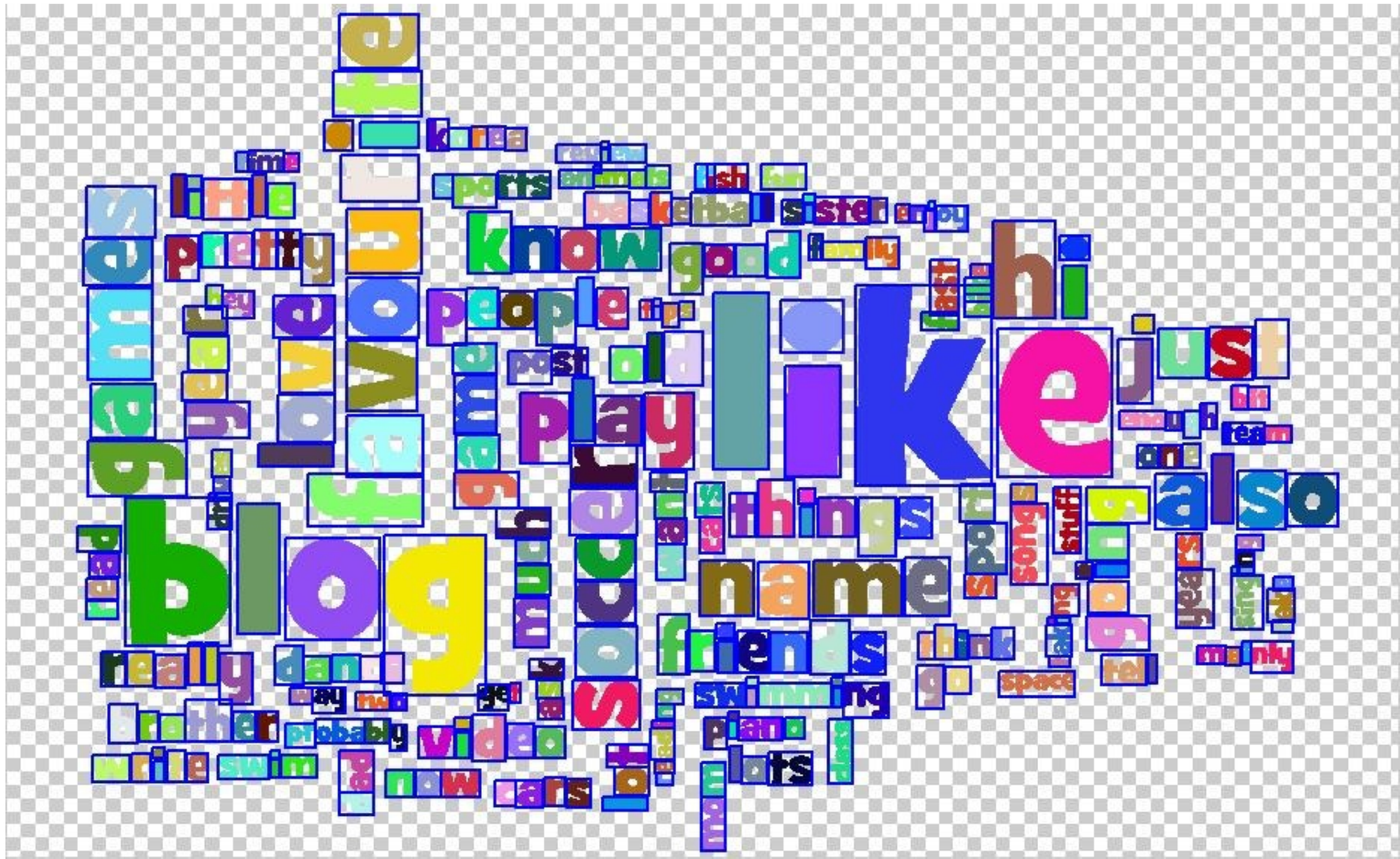} \\
  (b) \\
  \end{tabular}
  \caption{{The letters are extracted by finding the connected components of the image. (a) The extracted letters using connected component analysis. (b) The tight bounding boxes around the letters that are used for letter size estimation. }
  \label{fig:components}}
  \end{center}
\end{figure}

\subsection{Letter extraction}
In the first step we find the set of all letters which have appeared in the word cloud which is explained below. 

{\bf Finding the letter regions:} For recognizing the letters form a bitmap image we first need to find out which region in the image corresponds to a single letter to extract each and every  individual letter. This includes finding out which pixels correspond to a single letter. We assume that each letter has one connected segment in the image with the same color. The background has a different and distinctive color. We use a visual segmentation method for finding the region of each letter which is based on finding the connected components in an image. The connected component algorithm works by scanning an image pixel-by-pixel (from top to bottom and left to right) and group its pixels into components based on pixel connectivity, i.e. all pixels in a connected component share similar pixel intensity values and are in some way connected with each other. In Fig~\ref{fig:components} the extracted regions are shown for an input example. The colored letters in part (a) of the figure show the uniform regions that 
are discovered. In part (b) the tight bounding boxes around the extracted letters are shown. These boxes will be used for letter size estimation.

{\bf Assigning regions to letters:} Once the letter regions are extracted from the input image, we generate the foreground mask for each letter. This can be done by binarization of segmented regions into foreground/background. Then we create an image patch for each foreground mask and pass it to an Optical Character Recognition (OCR) algorithm. The output of this stage is the actual letters for each image patch. The OCR algorithm basically matches each input to its set of known letters and assigns the best match letter to each input patch.

\subsection{Word extraction}
After the word extraction is finished in the previous step, we will end up with a bag of disconnected letters. 
In order to construct the words out of the extracted letters, for each input word cloud we build a complete graph of all the detected letters and map the problem to a graph theoretic problem. For doing this, we assign each detected letter to a node in the graph. The edges of the graph are built based on the amount of similarity between the nodes.

{\bf Modeling nodes in the graph:} Each node $n$, has several properties which are defined based on the visual appearance of its corresponding letter. Basically, these properties are the spatial location $(x,y)$, color $c$ and the area $A$ of the tight bounding box around the letter defined using the width and height of the box. The node properties are important since they provide clues on to which word they belong.

{\bf Modeling edges in the graph:} Each detected letter is related to other letters in the graph based on the amount of similarity. Since the nodes have three properties of location, color and size, the similarity between any two nodes $n_i$ and $n_j$ will be defined based on the amount of agreement between these properties.

\begin{multline}
 W(n_i,n_j) = d_x(n_i,n_j) + d_y(n_i,n_j) + \\ d_{color}(n_i,n_j) + \\ d_{height}(n_i,n_j) + d_{width}(n_i,n_j)
\end{multline}

In the above equation $d_x$ and $d_y$ refer to the spatial distance in the $x$ and $y$ directions. Also, $d_{color}$ is the distance between in two colors in terms of RGB and $d_{height}$ and $d_{width}$ are the difference of the width and height of the two letters. 

\begin{figure}
  \begin{center}
  \includegraphics[scale=0.17]{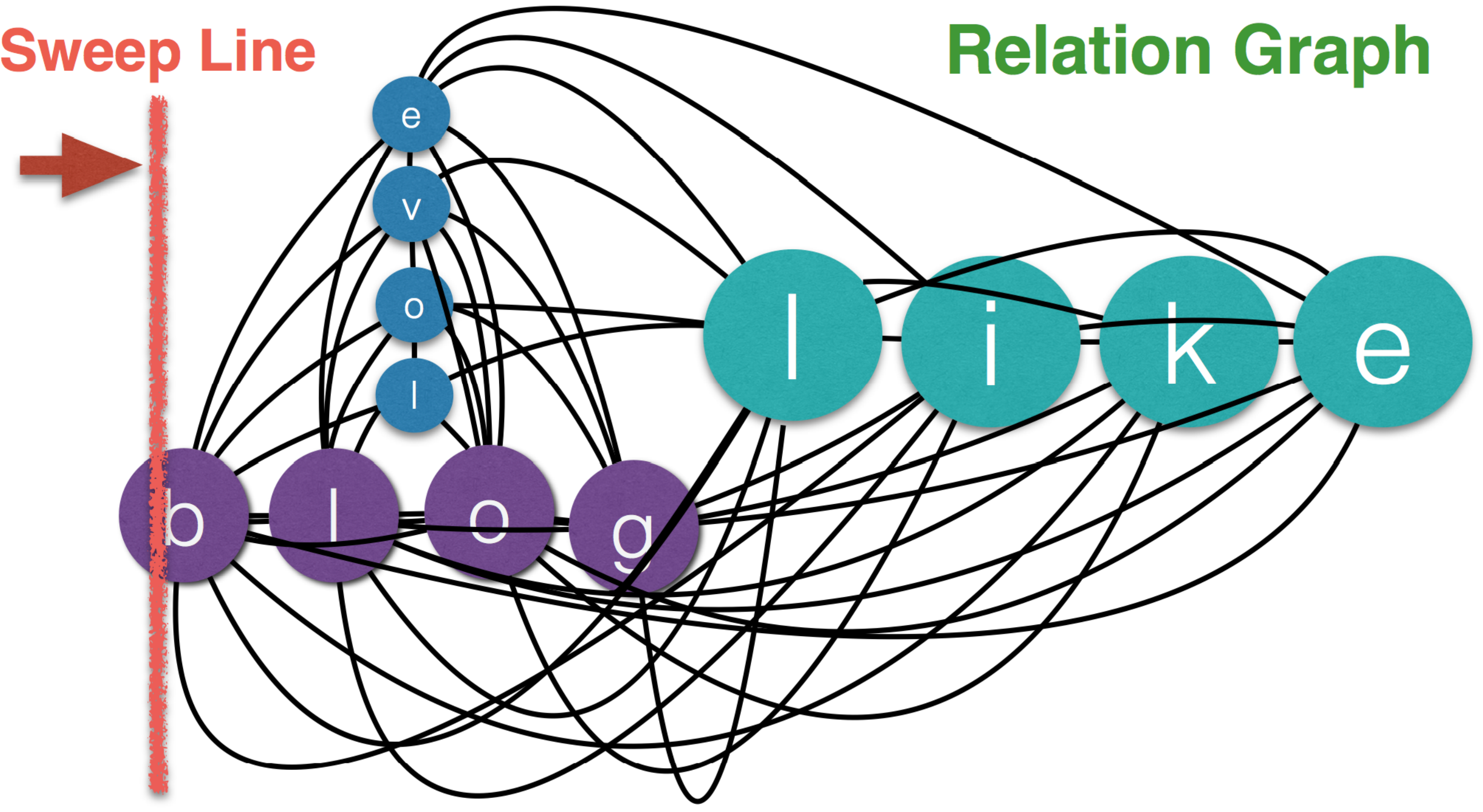}
  \caption{{The word cloud relation graph is a complete graph that connects all the nodes with a similarity weight. The sweep line visits the nodes from left to right to finds the words.}
  \label{fig:relgraph}}
  \end{center}
\end{figure}

{\bf Words as groups of similar letters:} After building the graph and defining the nodes and edges, the problem of word extraction will be reduced to finding groups of nodes with high similarity. Each of this groups will correspond to a word. For doing this, we find multiple cuts in graph and generate clusters of the nodes based on their similarities where each of these clusters will correspond to a word. Fig.~\ref{fig:relgraph} is an example which shows part of a graph for an input word cloud diagram.

{\bf Word construction using sweep line and bipartite graph matching:} 
For finding the words we start visiting each and every node in a sweep line fashion. For extracting the horizontal words, we move the sweep line from the left most side of the word cloud diagram to the right. At each time we move the sweep line $k$ pixels to the right and the sweep line will visit a number of nodes. Each of the visited nodes is a letter which corresponds to one of the current discovered words or will start making up a new word. In the beginning we start with no discovered words so the visited node in the first time slot will be considered as being the first letter of the words. For connecting the rest of the letters of a word to the currently discovered letters we construct a bipartite graph with the current visited nodes and their direct neighbors, in the following time slots. For finding the most confident matching letter, we solve a bipartite graph matching problem~\cite{clrs}. In this case, for each visited letter we find its adjacent letter by finding its best match (see Fig.~\ref{fig:
bipartite} 
for an example). After finding the best match for each of the visited nodes, we will remove the visited nodes and all the edges connected to them from the graph. We continue this, until we encounter a best matching edge whose weight is higher than a certain threshold $\tau$ we consider that edge as an invalid edge which means that we have visited all the letters of a word and it is completely extracted. This process continues until the whole image is sweeped by the sweep line. For extracting the vertical words, we repeat this procedure with a vertical sweep line where the sweep line will be moved from the bottom of the image toward the top.

\subsection{Size Estimation}
After finding the words, we compute the size of each word based on the area of the tight rectangular around the word divided by the number of letters in that word.
\begin{figure}
  \begin{center}
  \includegraphics[scale=0.17]{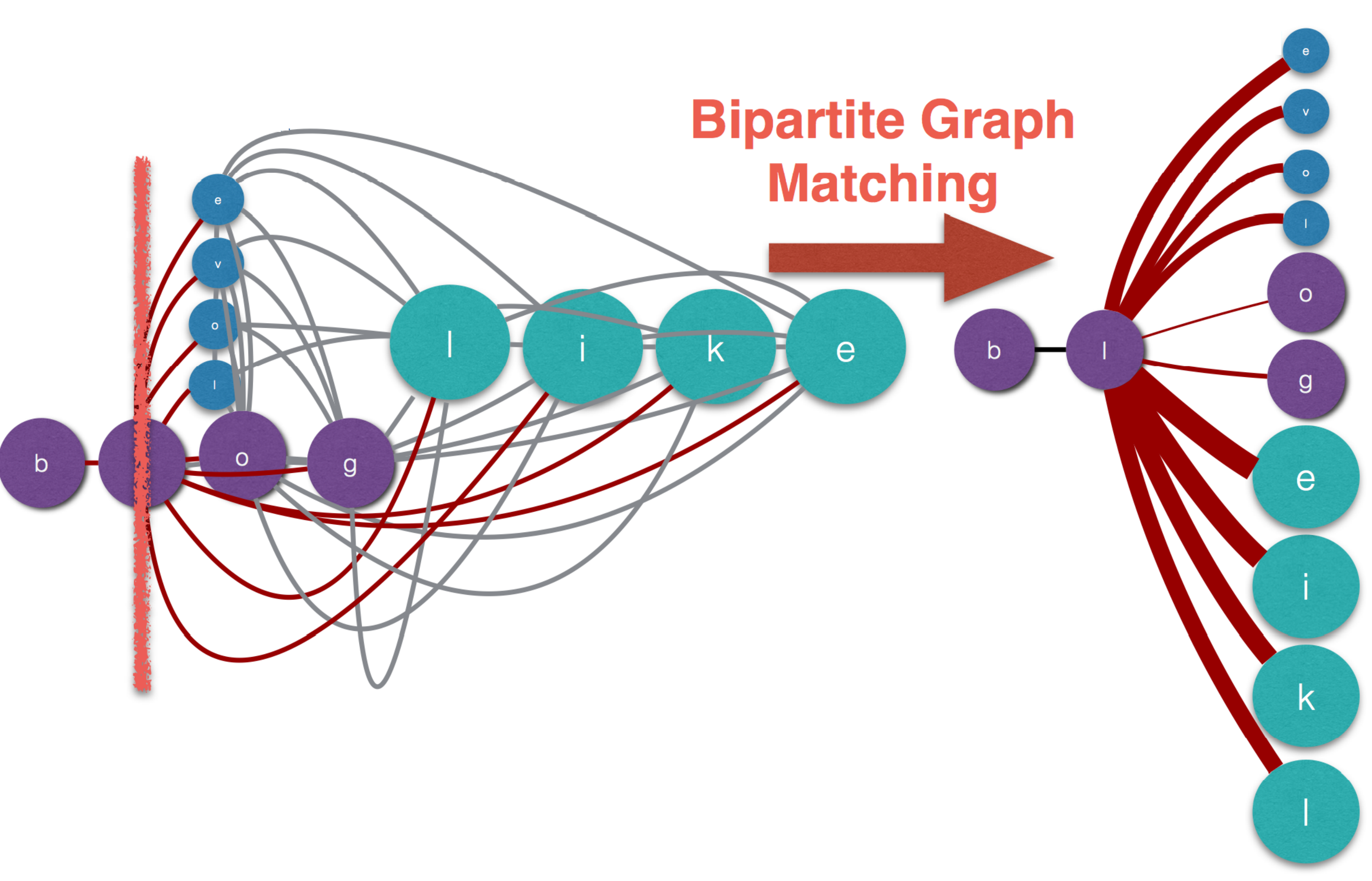}
  \caption{{An example which shows how our algorithm sweeps the input image and how we construct the bipartite graph at each time slot.}
  \label{fig:bipartite}}
  \end{center}
\end{figure}

\section{Results and Discussion}
For evaluating our method, we conducted two types of experiments. In the first part of the experiments we run our proposed algorithm on images collected from web and compare the results qualitatively. In the second part we produce word clouds using a visualization software and quantitatively evaluate the output by comparing against the ground truth data.

\subsection{Qualitative Experiments}
\begin{figure}
  \begin{center}
  \includegraphics[scale=0.5]{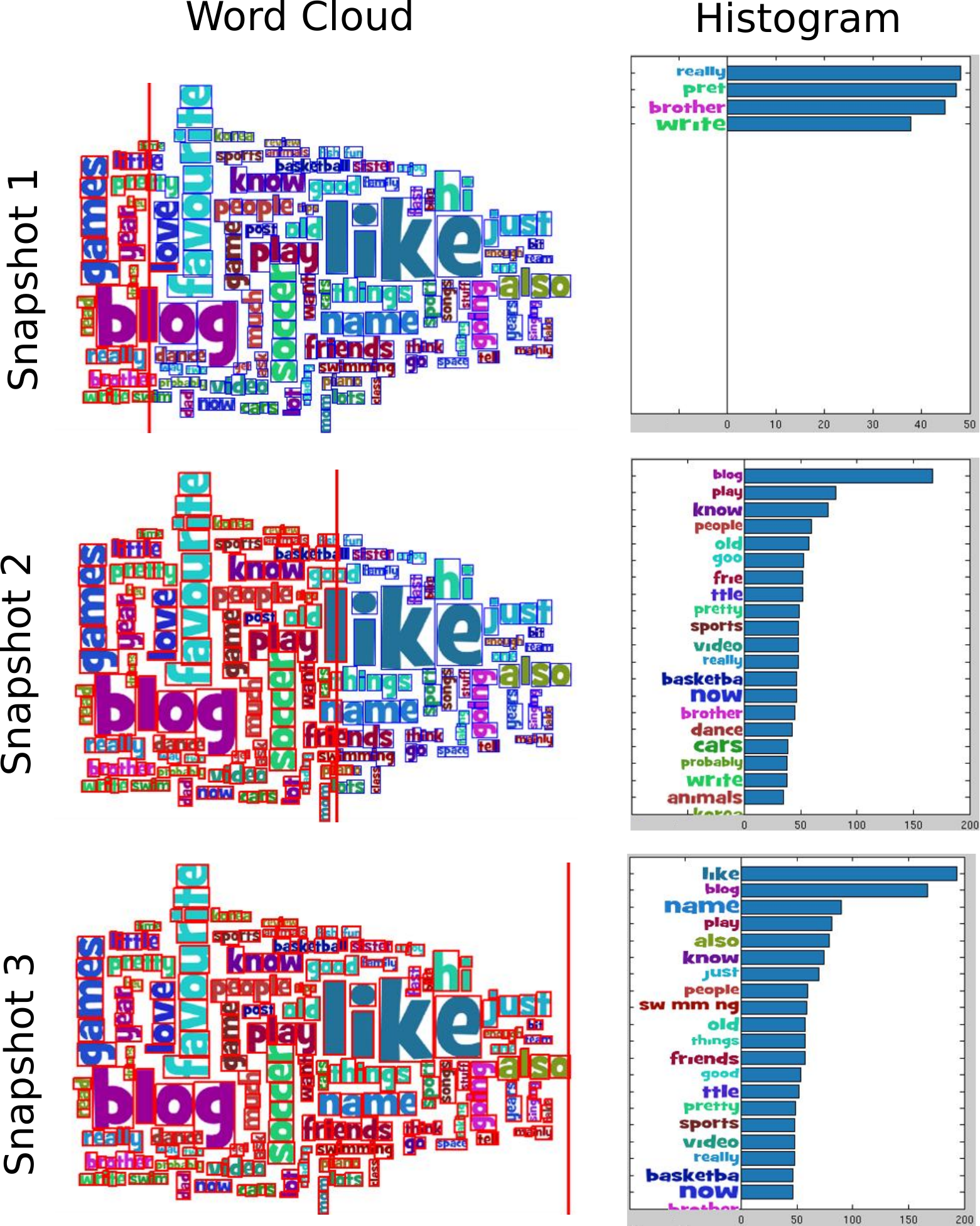}
  \caption{{An example of the output of our method in three random snapshots.}
  \label{fig:snapshot}}
  \end{center}
\end{figure}

We qualitatively evaluate the performance of our algorithm by running it on a collection of word clouds downloaded from Google images~\footnote{\url{https://www.google.com/search?site=imghp&tbm=isch&q=wordle}}. In this experiment we saved the results in different snapshots of the algorithm and produced the corresponding bar chart. These results are saved in the form of a video which visualizes the progress of our algorithm in finding the words and their weights. We observed that the produced histograms are consistent with the actual word clouds. An example of the snapshots of our algorithm running on a word cloud downloaded from Google is shown in Fig.~\ref{fig:snapshot}.

\subsection{Quantitative Experiments}
For conducting quantitative evaluation, we produce several word clouds from known raw data. For producing the word clouds we use the D3 implementation of wordle~\footnote{\url{http://www.jasondavies.com/wordcloud}}. In this case, we can have word clouds while we have access to their actual raw data. 
We selected several documents and used the aforementioned software to produce word clouds and saved the output in png and svg file formats. The png file is used for testing and the svg file is used for extracting the ground truth data. For generating the ground truth raw data, keywords along their weights are extracted from svg file. For quantitatively evaluating our results with the ground truth data, we compare our estimated size with the ground truth font size of the words. For each extracted word of our algorithm, we calculate its distance to the list of ground truth words and find its closest match since the OCR algorithm can have mistakes in predicting the right letters. We then compute the root mean square error (RMSE) of the predicted word sizes compared against the ground truth word size which will be the final error of our algorithm.

\begin{equation}
RMSE = \sqrt{\frac{\sum_{t = 1}^{N_{words}}{(S_{i}^{e} - S_{i}^{gt})^2}}{N_{words}}}
\end{equation}

\begin{figure}[t]
  \begin{center}
  \includegraphics[scale = .62]{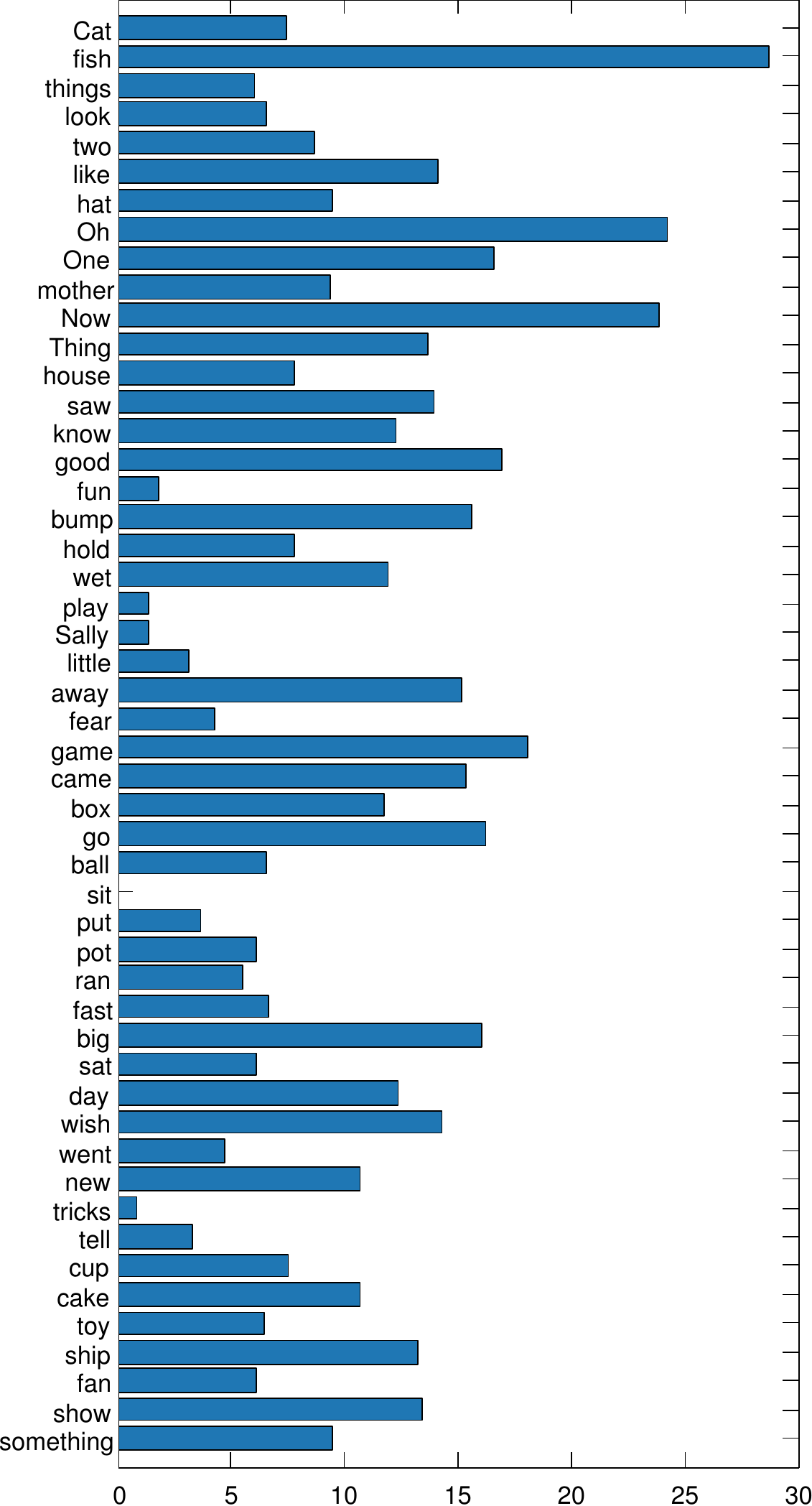}
  \caption{{Per word size estimation error in the D3 cloud document}
  \label{fig:d3cloud}}
  \end{center}
\end{figure}

In the above equation. $S_{i}^{e}$ and $S_{i}^{gt}$ refer to the estimated and ground truth size of the $i$th word and we have $N_{words}$ total number of words. 
Different types of documents are selected for doing the qualitative analysis which are as follows:
\begin{itemize}
 \item Text sample D3 word cloud (D3 cloud)
 \item Accepted conference papers (papers)
 \item CSE news (CSE)
 \item A news paper from New York Times (NYT)
 \item Wikipedia paper for interactive visualization (wiki)
\end{itemize}
Table \ref{table:res} summarizes the performance of our algorithm on the accuracy data extraction from the above list of the word clouds. We have provided the per word error size estimation error for the D3 cloud test set in Fig~\ref{fig:d3cloud}. To the best of our knowledge, this work has been the first attempt for extracting the raw information from word clouds so at this time there is no baseline to compare our results with. We believe that our algorithm has produced fairly good results both qualitatively and quantitatively. 

\begin{table}[t]
\begin{center}
\begin{tabular}{|c|c|c|c|c|c|}
\hline
& D3 cloud & papers & CSE & NYT & wiki \\
\hline
Error& 11.88 &14.19 & 25.44 & 14.09 & 9.62\\
\hline
\end{tabular}
\end{center}
\caption{\footnotesize{RMSE using five documents as the test set}}
\label{table:res}
\end{table}

\section{Conclusion and Future Work}
In this paper we proposed a novel method for automatically redesigning word clouds. Using our method we will be able to extract the raw data from a word cloud visualization. Our method is built upon extracting the letters using pixel wise connected component analysis and constructing words using graph theory methods. We have run several experiments to evaluate our method both quantitatively and qualitatively using downloaded word clouds from Google as well as self produced word clouds by D3. The results of our experiments show that our method is able to extract the words with their corresponding weight with considerably high accuracy. For further improving this method, we can use better OCR methods which incorporate probabilistic word completion techniques. Also, we can expand our current dataset to test the performance of our method more thoroughly.

%
%
%
%

{
\bibliographystyle{abbrv}
\bibliography{decoding}
}

\end{document}